\documentclass[sigconf, authorversion=true, nonacm=true]{acmart}
\usepackage{graphicx}
\usepackage{lipsum}
\usepackage{textcomp}
\usepackage{mathtools}
\usepackage{amsmath}
\usepackage{cuted}
\usepackage{footnote}
\begin{document}
\title{Feature Engineering for US State Legislative Hearings: Stance, Affiliation, Engagement and Absentees}
%
%


\author{Josh Grace}
\email{joshgrace15@gmail.com}

\affiliation{%
  \institution{California Polytechnic State University}
  \country{USA}
}
\author{Foaad Khosmood}
\email{foaad@calpoly.edu}

\affiliation{%
  \institution{California Polytechnic State University}
  \country{USA}
}
\begin{abstract}
In US State government legislatures, most of the activity occurs in committees made up of lawmakers discussing bills. When analyzing, classifying or summarizing these committee proceedings, some important features become broadly interesting. In this paper, we engineer four useful features, two applying to lawmakers (engagement and absence), and two to non-lawmakers (stance and affiliation). We propose a system to automatically track the affiliation of organizations in public comments and whether the organizational representative supports or opposes the bill. The model tracking affiliation achieves an F1 of 0.872 while the support determination has an F1 of 0.979. Additionally, a metric to compute legislator engagement and absenteeism is also proposed and as proof-of-concept, a list of the most and least engaged legislators over one full California legislative session is presented.

\end{abstract}


%
%
%
\maketitle              
\section{Introduction}
In representative democratic political systems, a legislature or parliament is the organization that creates laws and holds deliberations about them. US states, each being governed by a republican form of government, have state legislatures that produce laws. While much research is produced analyzing US Federal government and other parliamentary systems, relatively little work has been done for US State legislatures, where consequential debates, processes and relationships exist.

In this paper, we focus on four important features that are useful for statistical analysis of the legislature, and can be obtained from legislative committee transcripts. The first two are about non-legislator speeches. Non-legislators, such as invited experts and general public commentators, are an important part of the discourse that can shape legislation. When non-legislators speak, we would like to automatically extract their affiliations (if stated) and their stance on the legislation that is currently under discussion. We present models that can predict both with high levels of accuracy (F1 scores of 0.872 for affiliation and 0.979 for stance).

Two other features relate to legislators themselves: engagement and absenteeism. Legislator engagement is of particular importance to constituents as it can indicate their level of interest and attention to the legislation. While there's no universally accepted measure of engagement, we find that some factors such as asking questions and interacting with experts are often cited as proof of engagement. As our contribution for scholars, we propose an overall metric taking as variables several of these factors in consultation with an expert. Lastly, we also present a measure of absenteeism for lawmakers and discuss various challenges with this concept. 



\subsection{Motivation}
This works is part of a larger project called AI4Reporters\cite{ruprechter2018gaining} with the specific goal to help news reporters file more and better stories about the activities of the state legislature. The project hopes to fill a current gap in reporting as news organizations have significantly reduced their presence in state legislatures. The feature engineering and extraction that we discuss in this paper, along with other work, can provide automated tips to reporters and form the basis of a story or investigation. This will supplement the information available in AI4Reporters already and enable better tracking of how organizations’ comments on a bill can affect the outcome, and how specific organizations are influencing legislators. Furthermore, it will allow reporters to track how engaged specific legislators are in bill proceedings to ensure they are actively working to further their constituents’ interests and doing their part to improve the legislation under consideration. While previous work has been done to track organizations in bill proceedings and witness testimony, little research has been conducted on organizations influencing the legislature through public comments.

\subsection{Data}
We use high quality, human-verified transcripts from hearings in the California state legislature as developed by the Digital Democracy project~\cite{blakeslee2015digital} and already used in several other works~\cite{budhwar2018,kauffman2018,ruprechter2020,ruprechter2018}. The specific training set, engagement and affiliation statistics we use can be found online\footnote{https://iatpp.calpoly.edu/supporting-data-paper-feature-engineering-us-state-legislative-hearings-stance-affiliation}. We encourage reviewers to examine the data.

\subsection{Organization}
In the remainder of this paper, we examine some related work in section \ref{section:related}, followed by elaboration on our development process in section \ref{section:development}. In \ref{section:results} we present our results followed by Conclusion and Future work in sections \ref{section:conclusion} and \ref{section:future} respectively. 

\begin{figure}
 \includegraphics[width=\linewidth]{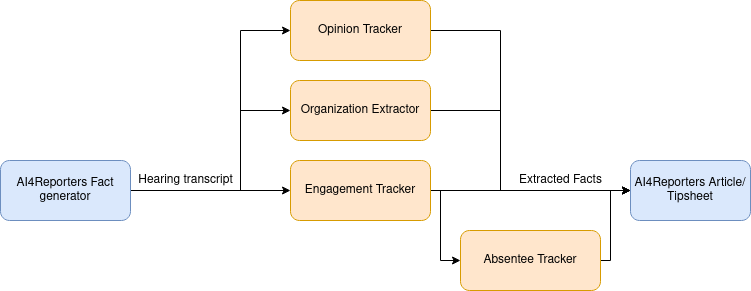}
 \caption{overview of article generation system}
\end{figure}

\section{Related Work}
\label{section:related}

While previous work had been done researching Named Entity Recognition (NER) in Congressional hearings, these taken a more general approach, interested in all named entities referenced during the proceedings. In this work however, NER is used to extract only organizations that are referenced as affiliated by the speaker. So modifications must be made to the system to ensure only these organizations are extracted. Additionally, while attempts have been made to track engagement in hearings, these have focused mostly on topic level engagement instead of engagement in the hearing. So, a very different approach is used to track this type of engagement.
\subsubsection{Legislative Discourse Analysis}
There have been a number of uses of NER in legislative discourse analysis. Dunham\cite{dunham2018dynamic} used NER to extract the organizations and political actors influencing US politics at the national and state level. This allows for tracking of names and titles of witnesses at congressional hearings and the organizational affiliations of these people. Dunham also disambiguated the organizations using NER to match organizations like NRA and National Rifle Association to the same entity. Angelidis et al. \cite{angelidis2018named} used NER to extract people, organizations, places, and legislative references in Greek legislative proceedings. Their work focused on extracting entities in Greek instead of English. Additionally, Koeva et al.\cite{koeva2020natural} did similar work in the Bulgarian legislature. They created a pipeline to extract named entities from the proceedings to annotate the documents by the referenced entities. The difference between this work and the previous entities referenced above is the type of entities desired. While these works were focused on extracting any entity mentioned in the hearing (other than Dunham who was interested in entities only in witness testimony), this work is interested specifically in affiliated organizations within public comments. This modification of the problem changes the requirement of the system, because a general NER system would have a high false positivity rate as it'd extract all entities, include those not affiliated with the speaker. So, the approaches used in previous work don’t generalize well in this application. Additionally, the format of public comments is fairly specific as many speakers use the same comment structure. So, this format can be leveraged to improve system performance.

\subsubsection{Legislative Engagement}
There has also been research into congressional engagement. Shafer\cite{shaffer2017cognitive} considered issue and topic engagement in Congressional hearings. By analyzing discourse around the 2008-2009 Financial Crisis, he showed that legislators engaged with a broader set of topics in hearings during the Crisis than before or after. This work is interested in engagement at the topic level, modelling the changes that occur when legislators interact with topics during and outside of crisis times. In contrast, the work described in this paper is interested in how engaged legislators are in a given hearing. 

More generally, discourse analysis has been applied to legislative proceedings with much success. Toft\cite{toft2010political} used discourse analysis to research how legislators justified welfare reform, which can be a very controversial subject in American politics. Kronick\cite{kronick2015rights} used a similar approach to analyze Canadian Parliamentary debates and the mandatory detention of migrant children. Davis\cite{davis1981description} examined committee hearing transcripts from 1971 to model legislator’s remarks during committee hearings on the bills under discussion. 

These works focus on modelling the deliberation around specific bills or controversial topics. But, the work in this paper focuses on creating a generalized analysis package that is applicable regardless of a specific bill or bill discussion.

\section{Development}
\label{section:development}
\subsubsection{Organizations Affiliated with Speakers}\label{organizational_affiliation_description}
To extract organization affiliation in public comments, a combination of the Stanford NER\cite{finkel2005incorporating}, SpaCy NER\cite{spacy}, and hand-crafted rules are used. To train the models, 693 comments from California Legislative hearings are manually tagged with the organizations that speakers stated affiliation to. Next, the original organizations are removed from the comments and replaced with a random organization from a list of 12,829 organizations registered with the California State Legislature. To train the Stanford NER model, for each of the 693 comments, 100 synthetic comments are generated with a random organization from the list of registered organizations. 

In total, this process yields 69,300 unique sentences to train the NER models on. These sentences are then tagged in the necessary format for each NER’s custom training interface. Instances of the speaker’s name are tagged with the person tag, the affiliated organizations are given the organization tag, and all other words are tagged with the ``other'' tag. By replacing the original organization in the training sentences, the model can be trained on significantly more data without additional effort in tagging data. Additionally, this style of training lets the models better learn the form of affiliation statements and common organizations giving public comment. 

To train the SpaCy NER model, a similar approach is used. Tagged comments are split by their affiliated organizations, but only comments with 0 or 1 affiliated organizations are used. Then, for each of the 12,829 registered organizations, 4 sentences with 1 affiliated organization are chosen and filled with the organization name itself. Then, the examples with 0 affiliated organizations are added to the training data. This gives 33,477 unique sentences to train on. 

The difference in training yields models with different uses. The SpaCy model is a high precision model. The organizations directly out of the model are highly likely to be true affiliates, but the model won’t capture all the affiliated organizations (low false positive, higher false negative rate). The Stanford NER model returns more organizations, if an organization is referenced as an affiliate in the comment, the Stanford NER model is highly likely to capture it. But, the Stanford NER model is more likely to include non-affiliated organizations as well (low false negative, higher false positive rate). This difference in performance is useful because the output can be combined for a low false positive and false negative rate.

To extract organizations from a hearing, the speech segments are first processed with the custom trained Stanford NER model detailed above. The model returns tuples of the word and tag. The output words are combined into the organizations outputted by Stanford NER and added to an organization list. Next, the groups in the known organizations are iterated over. If one of these organizations are present in the text but not in the output from Stanford NER, they’re added to the output list. Next, the input text is run with SpaCy NER and the output is combined in a similar way to the Stanford NER text into a second list of the output organizations referenced. The two organization lists are then compared, and if an organization is present in both, it’s added to the final output set. All unmatched outputs from SpaCy and Stanford NER are added to a separate set to determine what additional organizations should be added to the final output set. A rules based approach is used to determine which of the unmatched organizations should be included. Any organizations outside of the first 12 words (excluding tagged entities prior) of the sentence, are automatically ignored. This is effective as organizations are generally referenced at the start of a public comment as part of the introduction of the speaker. 

The organizations are also checked to ensure they aren’t identical to the speaker's name. Stanford NER sometimes returns the commenter’s name as an organization. This test ensures that the person’s name isn’t incorrectly included in the output. Next, the extracted organization is added to a test sentence and tagged with SpaCy NER to test if it is plausible as an affiliated organization with the more stringent model. If the entity passes all of these tests, it’s added to the output set. Finally, the outputs are checked to ensure they’re not perfect matches to the name of California cities or counties or ``board member'' and that they’re longer than 2 characters as these are also common errors in Stanford NER. The resulting set of output organizations is converted to a list and returned for inclusion in output sentences.

\subsubsection{Determining Commenter Opinion}
A rules based approach is used to track support or opposition in public comments. When giving comments, people generally use specific keywords to convey their opinion on bills. This includes phrases like ‘urge an aye vote’, ‘support this bill’, or ‘oppose this bill’. These keywords can be tracked to model the speaker’s position on a bill. When reading the text to determine support or opposition, these keywords are broken into 5 sets, strong opposition, strong support, medium opposition, medium support, and weak support. The phrases are listed in Table \ref{position_table}. 

The position tracker counts the number of occurrences of words from each of the five sets, and puts this vector into a decision tree classifier to determine if the commenter supports, opposes, or is neutral on the bill. While there are instances where public speakers support or oppose a bill, but don’t explicitly state this position, these instances were deemed sufficiently rare and treated as noise in the system. While tracking more subtle expressions of opinion would catch many of these instances, it would also significantly increase the risk of false positives, something the project strives to minimize for the sake of accurate reporting. When the speaker doesn’t explicitly state a position on the bill, they’re simply marked as neutral, which while not ideal, doesn’t risk false statements. Thus, an explicit, rules based approach was selected to reduce the rate of false positives.

\bgroup
\def\arraystretch{1.25}
\setlength{\tabcolsep}{12pt}
\begin{table*}
\caption{Position Categories and corresponding phrase lists}
\centering
\begin{tabular}{|c|c|} 
\hline
Category & Phrases\\
\hline\hline
Strong Opposition & ``oppose'', ``opposition'', ``opposing'', ``opposed''\\
\hline
Strong Support & ``support'', ``supporting''\\
\hline
Medium Opposition & ``no vote'', ``nay vote''\\
\hline
Medium Support & ``aye vote'', ``yes vote''\\
\hline
Weak Support & ``cosponsor''\\ 
\hline
\end{tabular}
\label{position_table}
\end{table*}
\egroup

\begin{equation}
\begin{aligned}
vote\_score_p={} \alpha * \frac{number\_votes_p}{num\_hearings\_on\_committee_p}
\end{aligned}
\label{vote_equation}
\end{equation}

\begin{equation}
\begin{aligned}
speaking\_score_p={} \beta * num\_times\_speaking_p
\end{aligned}
\label{speaking_equation}
\end{equation}

\begin{equation}
\begin{aligned}
back\_and\_forth\_score_p={} \gamma * num\_words\_in\_back\_and\_forth_p
\end{aligned}
\label{AoA_equation}
\end{equation}

\begin{equation}
\begin{aligned}
question\_score_p={} \delta * num\_questions_p
\end{aligned}
\label{question_equation}
\end{equation}

\begin{figure*}
\noindent
\begin{minipage}[t]{\textwidth}
\begin{equation}
\begin{aligned}[c]
engagement_p={} vote\_score_p + speaking\_score_p + back\_and\_forth\_score_p + question\_score_p
\label{engagement_equation}
\end{aligned}
\end{equation}
\end{minipage}%
\end{figure*}

\subsubsection{Quantifying Legislator Engagement}\label{legislator_engagement_description}
To track engagement, a rules based approach is applied using NLTK \cite{bird2009natural}. Engagement is modeled as detectable interaction within the proceedings through speaking during a hearing. This allowed for text based processing of a legislator’s dialog to track engagement. Four factors are considered to calculate engagement in the proceedings: verbally voting at the end of a hearing, speaking, back-and-forth conversations with non-legislators, and asking questions. These particular variables were chosen after discussions with the chief of staff to a former elected member of the leadership in the California State Senate. 

Verbal voting occurs at the end of a bill discussion, when the Committee Secretary calls each committee member's name for a vote. If a member responds to this call verbally, they're marked as voting for the hearing. Using this voting instead of final vote tally as published in the legislature website better tracks the Legislator's engagement in the hearing as they must be present live to respond to the voting request. 

The final vote score equation is detailed in equation \ref{vote_equation}, $\alpha$ is a scaling term for the final engagement equation, and the number of votes is divided by number of hearings to get a percent of hearings voted in. 

Each instance of speaking is defined as a roughly 30 second block of speaking by a legislator which is longer than 6 words. A Legislator speaking for 1 minute would be credited with 2 instances of speaking. The speaking score equation is detailed in equation \ref{speaking_equation}, $\beta$ is the scaling term. 

Back and forth conversations are instances where the legislator speaks, then a non-legislator speaks, then the same legislator speaks again (presumed reply). These are tracked in addition to raw speaking because they demonstrate a sustained interest in the hearing, instead of just a one-off comment or speech. This ensures that interactions with witnesses or bill presenters improves the legislator's score. Back and forths are filtered to ensure they have more than 12 words, and the number of words in all back and forths are summed per legislator. The back and forth score is generated in equation \ref{AoA_equation}, $\gamma$ is the scaling term for the back and forth. 

Finally, questions are counted using question marks as produced in the source transcript (which has been human verified). The transcripts include question marks at the end of each question, so this is a valid approximation. The number of question marks is summed for each legislator to find the number of questions asked. This is an interesting quantity as it shows the legislators were attentive enough in a hearing to ask for a clarification or further information from the witness or presenter. The question score is calculated in equation \ref{question_equation}, $\delta$ is a scaling term. 

The final engagement score is calculated by adding the previous 4 scores together and is detailed in equation \ref{engagement_equation}.

\subsubsection{Tracking Absent Legislators}
To track absenteeism, a rules based approach is used, similar to engagement tracking. 

To lay out the problem, there is no publicly available time table of when legislators are present in hearings. While face recognition from available videos is an option and has been implemented for other purposes, there is no guarantee that the cameras capture all sections of the room at all times. We believe the transcripts of hearings are the best data available to track absenteeism. 

We define a legislator as absent from a hearing if they don’t speak at all in the hearings (are not captured on the transcript) or vote verbally. It’s trivial to see that a legislator speaking in a hearing is sufficient evidence they were present in the proceedings, but showing that a legislator is absent is more difficult. To solve this, the vote roll call was used. At the end of most hearings, the committee secretary will call the legislator’s last name to request their vote, and generally will repeat that name to confirm the vote. If a legislator is absent from the hearing, there will be no voting response to the request. So, the committee secretary’s speech is parsed to determine if a given legislator was referenced when voting. In some instances, a bill isn’t voted on, and no discussion occurs other than the speaker or committee chair talking briefly. In these cases, the majority of the committee is marked as absent, despite the Legislator's speech being unnecessary. To handle this, if more than 60\% of the legislators on a committee are considered absent, we consider this a special meeting and no one is marked as absent in the final calculation. While this may miss some legislators who weren’t present at the proceedings, for the sake of reporting true facts, the impact of a false positive is higher than that of a false negative, so the system errs on the side of caution.

\section{Results}
\label{section:results}
\subsubsection{Organizational Extractor Testing Methodology}
To evaluate performance of the Organizational Extractor, the organizations are first extracted from the text using the Organizational Extractor. These entities are then compared to the true entities which were human annotated. When evaluating these matches, modification had to be made to accept minor deviations in the extracted organizations. For example, in the utterance ``Danielle Kendall Keiser on behalf of Common Sense Media, Common Sense Kids Action, in support'', the true organizations are ``Common Sense Media'' and ``Common Sense Kids Action'', but the Organizational Extractor returns ``Common Sense Media, Common Sense Kids Action''. In this case, the organizations are still correct and would be passable in an article text. So, the answer needs to be corrected to ensure the score is accurate. 

Organizations are converted to lowercase and are stripped of punctuation.  Then, the true and extracted organization lists are compared, and any differences are preserved for further analysis. Extracted organizations are then split by `and ’ and if ‘the ’ is a prefix, that is stripped too. These organizations are then compared again and used to calculate F1 scores. Unfortunately, these rules don’t capture all the mis-mappings possible. For example, in “Kisasi Brooks, with Policylink and the Alliance for Boys and Men of Color and I'm in strong support of this bill.”, the extracted organization is “Policylink and the Alliance for Boys and Men of Color”. While PolicyLink is correctly matched, “Alliance for Boys and Men of Color” isn’t. Any incorrect entities are manually reviewed to ensure they’re truly wrong. 

\subsubsection{SpaCy NER Model Performance}
In addition to the final training method for the SpaCy NER model described above, 4 other methods were considered for the model. The results of each model are included in Table \ref{spacy_ner_training_methods_performance_table}. The first model was trained on the hearings with their original organizations unchanged. This method had the worst performance, likely because there was little data to train the model on. Next, the model was trained on both the sentences with their original organizations and the organizations replaced. Third, the model was trained with all comments, regardless of the number of affiliated organizations. This likely had worse performance than the optimal method because the model didn't capture the tendency for affiliated organizations to occur earlier in the comment. Fourth, the model was trained by choosing 70 random organizations for each sentence and adding a sentence for each organization to the training data. Finally, the optimal method described previously was used. While it doesn't have the best false positivity or false negativity rate, it has the best trade off between the metrics (and highest F1). Thus, it was selected for use in the final model.

\bgroup
\setlength{\tabcolsep}{6pt}
\begin{table*}[h!]
\caption{Performance of SpaCy NER Training Methods}
\centering
\begin{tabular}{|c|c|c|c|c|} 
\hline
Training Method&True Positives&False Negatives&False Positives&F1\\
\hline\hline
Hearings with original organizations & 130 & 64 & 38 & 0.718\\
\hline
Hearings with organizations from list and original organization & 135 & 60 & 26 & 0.758\\
\hline
Hearings with organizations from list, regardless of number& 169 & 42 & 61 & 0.766\\
\hline
Hearing comments with 70 organizations& 136 & 65 & 20 & 0.762\\
\hline
Optimal method & 146 & 51 & 22 & 0.800\\
\hline
\end{tabular}
\label{spacy_ner_training_methods_performance_table}
\end{table*}
\egroup

\subsubsection{Organizational Extractor Model Performance}
For the final performance metrics, 3 models are considered. The optimal SpaCy NER model from above, the Stanford NER model, and the combined model using both Stanford NER and SpaCy NER. The metrics are included in Table \ref{organization_extactor_performance_table}. When training the Organizational Extractors, 693 examples were used, while 193 examples were used to test on. The table shows that the Stanford NER model is more likely to include an organization in a comment compared to SpaCy NER, while the SpaCy NER model is more certain than the Stanford NER model. But, when the models are combined, both False Negativity and False Positivity are reduced. So, the combination of the models ensures that the model can be both highly certain and include most referenced organizations.

\bgroup
\def\arraystretch{1.05}
\setlength{\tabcolsep}{12pt}
\begin{table*}[h!]
\caption{Performance of Organization Tracker variants}
\centering
\begin{tabular}{|c|c|c|c|c|} 
\hline
Model Type&True Positives&False Negatives&False Positives&F1\\
\hline\hline
Stanford NER & 149 & 45 & 47 & 0.764\\
\hline
SpaCy NER & 146 & 51 & 22 & 0.800\\
\hline
SpaCy NER and Stanford NER & 171 & 31 & 19 & 0.872\\
\hline
\end{tabular}
\label{organization_extactor_performance_table}
\end{table*}
\egroup

\bgroup
\setlength{\tabcolsep}{6pt}
\begin{table*}[h!]
\caption{Most Engaged California State Legislators}
\begin{tabular}{|l|c|c|c|c|c|c|}
\hline
Ranking&Legislator Name&Engagement Score&Voting Score&Speaking Score&Back and Forth Score&Question Score\\
\hline
\hline
1&Hannah-Beth Jackson&23.621&0.319&3.139&10.293&9.87\\
\hline
2&Mike McGuire&16.448&0.196&2.086&9.036&5.13\\
\hline
3&Benjamin Allen&13.44&0.235&1.79&4.875&6.54\\
\hline
4&Ricardo Lara&12.028&0.138&1.655&5.014&5.22\\
\hline
5&Jim Frazier&11.512&0.37&0.841&3.211&7.09\\
\hline
6&John Moorlach&11.34&0.331&1.082&2.906&7.02\\
\hline
7&Richard Pan&11.115&0.298&1.321&4.086&5.41\\
\hline
8&Nancy Skinner&10.604&0.261&1.758&4.144&4.44\\
\hline
9&Bob Wieckowski&10.553&0.29&1.296&3.887&5.08\\
\hline
10&Jim Beall&10.392&0.201&1.631&3.33&5.23\\
\hline
\end{tabular}
\label{top_engagement}
\end{table*}
\egroup

\bgroup
\setlength{\tabcolsep}{6pt}
\begin{table*}[h!]
\caption{Least Engaged California State Legislators}
\begin{tabular}{|l|c|c|c|c|c|c|}
\hline
Ranking&Legislator Name&Engagement Score&Voting Score&Speaking Score&Back and Forth Score&Question Score\\
\hline
\hline
113&Rob Bonta&0.507&0.081&0.071&0.114&0.24\\
\hline
114&Raul Bocanegra&0.486&0.086&0.083&0.077&0.24\\
\hline
115&Kevin Mullin&0.446&0.136&0.075&0.045&0.19\\
\hline
116&Ken Cooley&0.428&0.19&0.088&0.01&0.14\\
\hline
117&Sabrina Cervantes&0.422&0.304&0.023&0.015&0.08\\
\hline
118&Sydney Kamlager-Dove&0.289&0.069&0.02&0.04&0.16\\
\hline
119&Jimmy Gomez&0.276&0.036&0.026&0.054&0.16\\
\hline
120&Wendy Carrillo&0.219&0.086&0.067&0.016&0.05\\
\hline
121&Jesse Gabriel&0.092&0.076&0.005&0.011&0.0\\
\hline
122\footnotemark\footnotemark&Luz Rivas&0.068&0.047&0.004&0.008&0.01\\
\hline
\end{tabular}
\label{bottom_engagement}
\end{table*}
\egroup

\subsubsection{Position Tracker Model Performance}
Position tracking was also evaluated using F1 scores. The same dataset used for Organizational Extraction was used with annotations for the position of the speakers in their comments. 474 examples were used for training, while 167 examples were used in testing. The system had an F1 score of 0.9786 on the testing data.

\subsubsection{Most Common Affiliated Organization List}\label{common_affiliates_list}
The Organizational Affiliation system was applied to hearings from the California State Legislature 2017-2018 session to compute how engaged the legislators were over the session. The hearings start December 5th, 2016 and end August 31st, 2018 as legislative sessions last 2 years. From an initial list of 18,365 hearings from the session in the database, some hearings were filtered out to ensure quality data. First, hearings without votes were removed as the comments in these hearings aren't very applicable to the lawmaking process. Then, hearings without public comments were removed. After applying the filtering method, 5,182 hearings were processed. Using the method described in section \ref{organizational_affiliation_description}, the affiliated organizations were extracted from the public comments. 

These comments were added to a set for each hearing. Then, the affiliated organizations for each hearing were read and the occurrences of each organization were counted. Finally, highly frequent false positives were manually removed from the list, such as ``Members''. These organizations were then sorted by the number of hearings the occurred in. The number of hearings each organization commented in was counted instead of tracking the total number of comments from an organization as certain groups will give numerous comments in a small number of hearings, which will skew the counts towards these organizations. This work was interested in groups with a broader influence over many issues. So, the number of hearings commented on was preferred. There were 39,258 total organizations referenced in hearings and 15,334 unique organizations, but this includes False Positives and variations on the same group name, like ``SEIU'' v.s. ``Service Employees International Union''. The 10 organizations commenting on the most hearings are listed in Table \ref{top_affiliates_commenting}.

\bgroup
\setlength{\tabcolsep}{6pt}
\begin{table}[h!]
\caption{Organizations with Affiliates Commenting on the Most Hearings}
\begin{tabular}{|l|c|}
\hline
Ranking&Organization Name\\
\hline
\hline
1&ACLU of California\\
\hline
2&California Labor Federation\\
\hline
3&Western Center on Law and Poverty\\
\hline
4&Sierra Club California\\
\hline
5&California Chamber of Commerce\\
\hline
6&California State Association of Counties\\
\hline
7&League of California Cities\\
\hline
8&California Federation of Teachers\\
\hline
9&California District Attorneys Association\\
\hline
10&State Building and Construction Trades Council\\
\hline
\end{tabular}
\label{top_affiliates_commenting}
\end{table}
\egroup

\subsubsection{Legislator Engagement Constants}
After interviewing an expert, a former chief of staff to a California legislator and party leader,  asking questions and engaging in back and forth conversations were identified as the most important engagement metrics, while Speaking time alone was identified less important, and voice voting was considered least impactful. Constants were chosen to reflect the expert analysis and are listed in table \ref{engagement_constants}. $\beta$ and $\gamma$ are especially small as the back and forth word count and speaking counts can get quite large. The constants need to be small to keep the values in a similar range.

\bgroup
\setlength{\tabcolsep}{6pt}
\begin{table}[h!]
\caption{Engagement Constants}
\begin{tabular}{|l|c|}
\hline
Name&Value\\
\hline
\hline
$\alpha$&0.5\\
\hline
$\beta$&0.0005\\
\hline
$\gamma$&0.00005\\
\hline
$\delta$&0.01\\
\hline
\end{tabular}
\label{engagement_constants}
\end{table}
\egroup

\subsubsection{Legislator Engagement List}
The Legislative Engagement Modeling system was applied to find legislator engagement rankings with a similar process to section \ref{common_affiliates_list}. The hearings were from the same California State legislative session database and time period. 

First, hearings without votes were removed as the discussions in these hearings are likely not substantive because they had no legislative outcome. Then, hearings with less than 3 legislators speaking were removed as many hearing are simply procedural and feature only the chairperson or an additional speaker, which would skew the rankings in favor of chairs. Additionally, floor hearings were removed as all legislators aren't expected to speak during the hearing, so instances of speaking is a poor proxy for engagement in this hearing type. 

After filtering hearings, 4,949 hearings were processed. Using the method described in section \ref{legislator_engagement_description}, each legislator had their engagement score with the data in the hearings they were on the committee for. The 10 most engaged Legislators are detailed in Table \ref{top_engagement}. Additionally, the 10 least engaged Legislators are listed in Table \ref{bottom_engagement}.

\footnotetext[1]{Anthony Rendon and Kevin De Leon are technically the least engaged legislators as they aren't members of any voting, non-floor session committees. But, they were excluded from the list because we were unable to collect representative statistics for them.}
\footnotetext[2]{Legislative leadership can have low engagement scores because they aren't as engaged in committee hearings, and instead work outside the public hearings}

\section{Conclusion}
\label{section:conclusion}
Using NER, a system to extract affiliated organizations in legislative public comments is proposed with an F1 score of 0.872. This system is applied to the 2017-2018 California State legislative session to extract a list of the groups mostly often using the public comments to interact with the legislative process. A system is also introduced to track the position of the Public Commenters, which has an F1 score of 0.979.

Additionally, a method to quantify legislator engagement in hearings is proposed. It incorporates features identified by an authority on the California State legislature to be meaningful to the legislator's engagement in a hearing. 

Furthermore, members shown to be highly engaged in table \ref{top_engagement} were confirmed to be generally accurate by the same expert interviewed by the authors. Similarly, a method to track absenteeism was proposed that quantifies how frequently legislators miss their committee hearings. These introduce novel methods for quantifying how active legislators are in their committee hearings, which can inform the general public in future legislative elections.

\section{Future Work}
\label{section:future}
With the addition of organization tracking within hearings, work can be done to analyze organizations between hearings. Various organizations, particularly lobbying groups give comments on many bills. Accordingly, it would be interesting to track these comments over time, and evaluate how their positions on different bills change. This would give insight into the goals of these organizations and would shed a light on how these private groups influence the legislative process. Furthermore, some groups often lobby together, such as special interest groups representing the same industry, so it would be interesting to track organizations that have high correlation in their position on bills considered by the legislature. As groups giving public comment on bills can be tracked between hearings, work should be done to mine this data and find associations and trends within this data.

The engagement score tracking could be extended to other state legislatures. As the method only uses hearings transcripts, it is applicable to legislatures that produce video or written records of the proceedings. Furthermore, the method doesn't rely on any California State legislature specific features that would preclude its use in other state legislatures. So, additional work should extend this work to track engagement across more state legislatures. With engagement tracked across many state legislatures, it would be interesting to make comparisons in legislator engagement between the legislatures to quantify how closely external groups can view the Legislative process. Furthermore, it would be interesting to account for committee position when calculating engagement. Chairpersons of a committee would be expected to interact more with the proceedings than a normal member, which the proposed engagement tracking doesn't account for.

\section*{Acknowledgment}
The authors thank the Cal Poly Institute for Advanced Technology and Public Policy and Ms. Christine Robertson for their valuable insight and support of this project.

%
%
%
{\large
\bibliographystyle{refs-style}
\bibliography{refs}}

\begin{thebibliography}{10}
\providecommand{\url}[1]{\texttt{#1}}
\providecommand{\urlprefix}{URL }
\providecommand{\doi}[1]{https://doi.org/#1}

\bibitem{angelidis2018named}
Angelidis, I., Chalkidis, I., Koubarakis, M.: Named entity recognition, linking
  and generation for greek legislation. In: Palmirani, M. (ed.) Legal Knowledge
  and Information Systems - {JURIX} 2018: The Thirty-first Annual Conference,
  Groningen, The Netherlands, 12-14 December 2018. Frontiers in Artificial
  Intelligence and Applications, vol.~313, pp. 1--10. {IOS} Press (2018).
  \doi{10.3233/978-1-61499-935-5-1},
  \url{https://doi.org/10.3233/978-1-61499-935-5-1}

\bibitem{bird2009natural}
Bird, S., Klein, E., Loper, E.: Natural language processing with Python:
  analyzing text with the natural language toolkit. " O'Reilly Media, Inc."
  (2009)

\bibitem{blakeslee2015digital}
Blakeslee, S., Dekhtyar, A., Khosmood, F., Kurfess, F., Kuboi, T., Poschman,
  H., Prinzivalli, G., Robertson, C., Durst, S.: Digital democracy project:
  Making government more transparent one video at a time. Digital Humanities
  2015  (2015)

\bibitem{budhwar2018}
Budhwar, A., Kuboi, T., Dekhtyar, A., Khosmood, F.: Predicting the vote using
  legislative speech. In: Proceedings of the 19th annual international
  conference on digital government research: governance in the data age. pp.
  1--10 (2018)

\bibitem{davis1981description}
Davis, K.M.: A description and analysis of the legislative committee hearing.
  Western Journal of Speech Communication  \textbf{45}(1),  88--106 (1981)

\bibitem{dunham2018dynamic}
Dunham, J.J.W.: Dynamic responsiveness in the American states: legislators,
  constituents, and organized interests; and, Do Legislators Respond to
  Redistricting?: positioning in the California Legislature. Ph.D. thesis,
  Massachusetts Institute of Technology (2018)

\bibitem{finkel2005incorporating}
Finkel, J.R., Grenager, T., Manning, C.D.: Incorporating non-local information
  into information extraction systems by gibbs sampling. In: Proceedings of the
  43rd Annual Meeting of the Association for Computational Linguistics
  (ACL’05). pp. 363--370 (2005)

\bibitem{spacy}
Honnibal, M., Montani, I., Van~Landeghem, S., Boyd, A.: spacy:
  Industrial-strength natural language processing in python (2020).
  \doi{10.5281/zenodo.1212303}, \url{https://doi.org/10.5281/zenodo.1212303}

\bibitem{kauffman2018}
Kauffman, D., Khosmood, F., Kuboi, T., Dekhtyar, A.: Learning alignments from
  legislative discourse. In: Proceedings of the 19th annual international
  conference on digital government research: Governance in the data age.
  pp.~1--2 (2018)

\bibitem{koeva2020natural}
Koeva, S., Obreshkov, N., Yalamov, M.: Natural language processing pipeline to
  annotate bulgarian legislative documents. In: Proceedings of The 12th
  Language Resources and Evaluation Conference. pp. 6988--6994 (2020)

\bibitem{kronick2015rights}
Kronick, R., Rousseau, C.: Rights, compassion and invisible children: A
  critical discourse analysis of the parliamentary debates on the mandatory
  detention of migrant children in canada. Journal of Refugee Studies
  \textbf{28}(4),  544--569 (2015)

\bibitem{ruprechter2020}
Ruprechter, T., Khosmood, F., G{\"u}tl, C.: Deconstructing human-assisted video
  transcription and annotation for legislative proceedings. Digital Government:
  Research and Practice  \textbf{1}(3),  1--24 (2020)

\bibitem{ruprechter2018gaining}
Ruprechter, T., Khosmood, F., Kuboi, T., Dekhtyar, A., G{\"u}tl, C.: Gaining
  efficiency in human assisted transcription and speech annotation in
  legislative proceedings. In: Proceedings of the 19th Annual International
  Conference on Digital Government Research: Governance in the Data Age.
  pp.~1--2 (2018)

\bibitem{ruprechter2018}
Ruprechter, T., Khosmood, F., Kuboi, T., Dekhtyar, A., G{\"u}tl, C.: Gaining
  efficiency in human assisted transcription and speech annotation in
  legislative proceedings. In: Proceedings of the 19th Annual International
  Conference on Digital Government Research: Governance in the Data Age.
  pp.~1--2 (2018)

\bibitem{shaffer2017cognitive}
Shaffer, R.: Cognitive load and issue engagement in congressional discourse.
  Cognitive Systems Research  \textbf{44},  89--99 (2017)

\bibitem{toft2010political}
Toft, J.: The political act of public talk: How legislators justified welfare
  reform. Social Service Review  \textbf{84}(4),  563--596 (2010)

\end{thebibliography}
\end{document}